\documentstyle[prl,aps]{revtex}

\input psfig
\begin{document}
\def\vev#1{{\langle#1\rangle}}
\def\lsim{\mathrel{\raise.3ex\hbox{$<$\kern-.75em\lower1ex\hbox{$\sim$}}}}
\draft
\title{How accurately can one test CPT conservation with reactor and solar neutrino
experiments?}
\author{$^1$John N. Bahcall\thanks{E-mail: jnb@sns.ias.edu}, 
$^2$V. Barger\thanks{E-mail: barger@pheno.physics.wisc.edu} and 
$^3$Danny Marfatia\thanks{E-mail: marfatia@physics.bu.edu}}
\vskip 0.3in
\address{$^1$Institute for Advanced Study, Princeton, NJ 08540}
\vskip 0.1in
\address{$^2$Department of Physics, University of Wisconsin-Madison, WI 53706}
\vskip 0.1in
\address{$^3$Department of Physics, Boston University, Boston, MA 02215}
\maketitle
\begin{abstract}
We show that the combined data from solar neutrino experiments and
from the KamLAND reactor neutrino experiment can establish an upper
limit on, or detect, potential CPT violation in the neutrino sector of
order $10^{-20}\, {\rm GeV}$ to $10^{-21} \, {\rm GeV}$. 
\end{abstract}


A number of previous authors have discussed the possibility of
observing hypothesized CPT violation through neutrino oscillation
phenomena~\cite{prescient,coleman,barger,barenboim,murayama}.
Stimulated by this fascinating possibility, we investigate here a
related question: How accurately can one use solar and reactor
neutrino measurements to set an upper limit on CPT violation in the
neutrino sector? In this paper, therefore, we assume CPT violation is
small and determine the upper limit that can be set on the magnitude
of possible violations of CPT using existing solar neutrino data and
reactor neutrino data that will soon be available~\cite{piepke}. Of
course, CPT violation may be detected if differences between neutrino
and anti-neutrino propagation are observed that are larger than the
sensitivity limit derived here.

Solar neutrino experiments give information about the propagation
characteristics of neutrinos, primarily concerning the two mass
eigenstates with the smallest absolute difference in their
masses. Reactor experiments, such as KamLAND~\cite{piepke}, can give
similar information about anti-neutrinos. Similar particle energies
(1-10 MeV) are involved in both sets of experiments.

If CPT is conserved, then the survival probabilities  as a
function of energy, P(E), must be identical in vacuum for neutrinos and for
anti-neutrinos. The reader will easily recognize that the identity of
the survival probabilities for neutrinos and anti-neutrinos is a
special case of the general result of  CPT conservation:
$P_{\alpha,\beta} = P_{\bar{\beta},\bar{\alpha}}$.

We shall show below that solar neutrino
experiments~\cite{solarexperiments} plus the reactor experiment
KamLAND~\cite{piepke} will, taken together, be sufficiently accurate
to test significantly the equality of survival probabilities and, if
CPT is conserved, to set a stringent upper limit on some conjectured
forms of the violation.

First, we want to make some order of magnitude comparisons in
order to motivate the fact that the KamLAND/solar neutrino data
are precise enough to give interesting limits on CPT. The best
established limit on CPT violation in the baryon sector is the
well-known upper limit on the mass difference between K and
$\bar{{\rm K}}$~\cite{pdg}:
\begin{equation} |m_{\rm K} - m_{\rm
\bar{K}}|~<~4.4\times10^{-19}\, {\rm GeV} \ \ \ \ (90\%\ {\rm{C.L.}}).
\label{eq:kmassdifference}
\end{equation}

How does this upper limit on the K-$\bar{{\rm K}}$ mass difference
compare with the characteristic range of sensitivity of solar
neutrino and reactor neutrino experiments?

Two energy parameters naturally affect the sensitivity of solar
and reactor neutrino probes of CPT violation. The first
parameter is just the time available for rare processes to occur,
which is inversely proportional to the separation, $L$, between
the source of the reactor neutrinos  and the detector. The energy
scale which results from this consideration is
\begin{equation} \delta CPT ~\sim~\hbar c/L ~\sim~10^{-21}({\rm 200~ 
km/L})\,{\rm
GeV}, \label{eq:deltacpthcl}
\end{equation}
where for specificity we have used the characteristic reactor-detector
distance that applies to KamLAND, $L \sim 200\,$km. The sensitivity
also depends upon the frequency of the neutrino oscillations, which is
determined by the neutrino energy and the difference of squared
neutrino masses. Thus we have a second parameter affecting the
sensitivity of CPT measurements involving solar neutrinos,
\begin{equation}
\delta CPT ~\sim~ \delta m^2/E ~\lsim~10^{-19}\,{\rm GeV}.
\label{eq:deltacptmass}
\end{equation}
In obtaining the numerical form of Eq.~(\ref{eq:deltacptmass}), we
used $\delta m^2 < 4\times 10^{-4}$ eV$^2$ for solar neutrino
oscillations~\cite{jnb}.

 Comparing the dimensional estimates
of $\delta CPT$ from Eq.~(\ref{eq:deltacpthcl}) and
(\ref{eq:deltacptmass}) with the K-$\bar{{\rm K}}$ mass difference, we
see that solar and reactor neutrino observations can indeed set a
sensitive upper limit on (or perhaps measure) CPT violation. The
$\delta CPT$ sensitivity for solar and reactor neutrinos is
expected to be one or two orders of magnitude more sensitive than
the existing upper limit to the K-$\bar{{\rm K}}$ mass difference.

The basic strategy we use in evaluating the sensitivity of solar
neutrino and reactor experiments to CPT violation is to first suppose
that CPT is exactly conserved. Then we find the maximum difference
between the allowed survival probabilities for neutrinos and
anti-neutrinos that is consistent with the expected experimental
uncertainties. To be conservative and specific, we adopt for solar
neutrinos the currently allowed region (in $\delta m_\nu^2$ and
$\sin^2{2\theta_{{\nu}}}$ space) permitted by existing
experiments~\cite{jnb} (see also~\cite{seealso}). For the KamLAND
experiment, we assume that the parameter space that will be found for
anti-neutrino oscillations is a fraction~\cite{BMW} $\epsilon \leq 1$
of the allowed solar neutrino parameter space. We assume that the
correct solar neutrino solution is the favored LMA solution, since if
CPT is conserved KamLAND will be sensitive to anti-neutrino
oscillations only if the LMA solution has been chosen by nature.

We characterize the general sensitivity of reactor and solar
neutrino experiments to CPT violation by the quantity 
\begin{equation}
\vev{\Delta CPT} ~=~2 \frac{|\vev{\rm P_{\nu \nu}(E,L) -
P_{\bar{\nu}\bar{\nu}}(E,L)}|} {\vev{{\rm P_{\nu \nu}(E,L) +
P_{\bar{\nu}\bar{\nu}}(E,L)}}}. \label{eq:defndeltacpt}
\end{equation}
Here both ${\rm P_{\nu \nu}(E,L)}$ and ${\rm
P_{\bar{\nu}\bar{\nu}}(E,L)}$ are computed for the same experimental
situation, but with different values for ($\delta m^2_\nu$, $\sin^2
2\theta_\nu$) and for ($\delta m^2_{\bar{\nu}}$, $\sin^2
2\theta_{\bar{\nu}}$).  The average in Eq.~(\ref{eq:defndeltacpt}) is
carried out over reactor distances $L$ and over all neutrino and
anti-neutrino energies $E$.

An experimental upper limit on $\vev{\Delta CPT} $ can be
used to test arbitrary future conjectures of CPT violation. Practically 
speaking,
 $\vev{\Delta CPT}$ is the number of
events observed in a reactor (anti-neutrino) oscillation minus the
number of events that would have been observed if neutrinos and
anti-neutrinos had exactly the same oscillation parameters,
divided by the average number of neutrino and anti-neutrino
events.

We have evaluated numerically the maximum value of $\vev{\Delta CPT}$
 that results from the mismatch of the average
survival probabilities computed from two different points within
the same allowed neutrino and anti-neutrino oscillation regions,
both of which are assumed coincident with the current $3\sigma$
allowed parameter domain for solar neutrinos ($\epsilon = 1$). We
find for $10^6$ randomly sampled pairs,
\begin{equation}
\vev{\Delta CPT} ~\leq~1.1\ \ \ \ (3\sigma~{\rm limit}).
 \label{eq:DeltaCPTupper}
\end{equation}
The maximum value is achieved for the pairs of neutrino parameters,
$\delta m_\nu^2 = 3.3\times10^{-5}\, {\rm eV^2}$,
$\sin^2{2\theta_{{\nu}}} =0.98$; $|\delta m_{\bar{\nu}}^2| = 6.4\times
10^{-5}\, {\rm eV^2}$, $\sin^2{2\theta_{\bar{\nu}}=0.6}$, which lie
near the boundary of the currently allowed solar neutrino
oscillation region.  Matter effects that simulate CPT violation,
and which arise from the different interaction strengths of neutrinos
and anti-neutrinos with electrons in the earth, would contribute 
 $\sim 0.1$ to $\vev{\Delta CPT}$.

We have tested the sensitivity of the upper limit given in
Eq.~(\ref{eq:DeltaCPTupper}) to the assumed size, characterized by
$\epsilon$, of the allowed parameter domain for the KamLAND
experiment. Even choosing $\epsilon$ as small as $0.25$ does not
appreciably change the
upper limit that is obtainable for $\vev{\Delta CPT}$.
Note that this bound applies for any source of CPT violation,
Lorentz symmetry preserving or violating.

To illustrate the power of the combined KamLAND reactor and solar
neutrino experiments, we consider an effective interaction which
has been discussed by Coleman and Glashow~\cite{coleman},
and by Colladay and Kostelecky~\cite{colladay}, 
that violates both Lorentz invariance and CPT
invariance. The interaction is of the form
\begin{equation}
{\mathcal{L}}(\Delta CPT)) ~=~\bar{\nu}_{\rm
L}^\alpha{b^{\alpha\beta}_\mu}{\gamma_\mu}{\nu_{\rm L}^\beta},
 \label{eq:interaction}
\end{equation}
where $\alpha, \beta$ are flavor indices, L indicates
that the neutrinos are left-handed, and $b$ is a Hermitian matrix. We discuss 
the
special case with rotational invariance in 
which $b_0$ and the mass-squared matrix are diagonalized
by the same mixing angle. We also assume that there are only two
interacting neutrinos (or anti-neutrinos) and follow previous
authors in defining $\eta$ as the difference of the phases in the
unitary matrices that diagonalize $\delta m^2$ and the CPT 
odd quantity $\delta b$, which is the difference between
the two eigenvalues of $b_0$. 

When the relative phase $\eta =0$, the expression for the
survival probabilities of neutrino and anti-neutrinos take on an
especially simple form, see~\cite{barger}.   We find
the upper limit that could be established for $\delta b$ if the
allowed anti-neutrino oscillation region determined by KamLAND
were equal to the current allowed solar neutrino oscillation
region. In agreement with an approximate analytic solution for
this case of the form given in Eq.~(\ref{eq:deltacpthcl}), we
find by a numerical exploration that
\begin{equation}
\delta b ~<~1.6 \times 10^{-21}\,{\rm GeV} , ~~\eta = 0\ \ \ \ (3\sigma~{\rm limit}).
\label{eq:deltabeta0}
\end{equation}
It will be difficult to improve this limit by an order of magnitude
because earth matter effects, which simulate CPT violation, are
comparable to intrinsic CPT violation for
\begin{equation}
\delta b \sim \sqrt{2} G_F N_e \approx 1.5\times 10^{-22}\,{\rm GeV},  
 \label{eq:matter}
\end{equation}
where $N_e$ is the electron number density in the earth's crust.

If the relative phase $\eta = \pi/4$, the neutrino masses and the
CPT violating term appear in the expressions for the survival
probabilities in the form [$(\delta m^2/E)^2 + (2 \delta b)^2$].
We again explore  numerically an assumed KamLAND allowed region
equal to the current allowed region for solar neutrino
oscillations. We find
\begin{equation}
\delta b ~<~3.1 \times 10^{-20}\, {\rm GeV}, ~~\eta =
\pi/2\ \ \ \ (3\sigma~{\rm limit}),
\label{eq:deltabetapiovertwo}
\end{equation}
in good agreement with the dimensional estimate given in
Eq.~(\ref{eq:deltacptmass}). 
The limit in Eq.~(\ref{eq:deltabetapiovertwo}) could be improved 
by about a factor of three if solar neutrino experiments exclude
$\delta m_{\nu}^2>10^{-4}$ eV$^2$ and KamLAND also finds 
$|\delta m_{\bar{\nu}}^2|<10^{-4}$ eV$^2$. 

We conclude that the combined data from solar and reactor neutrino
experiments can test  accurately  the conservation of CPT in the
neutrino sector. If a reactor experiment like KamLAND finds
anti-neutrino survival probabilities outside the range expected for
neutrinos (based on solar neutrino experiments,
cf. Eqs.~\ref{eq:DeltaCPTupper},\ref{eq:deltabeta0}, and
\ref{eq:deltabetapiovertwo}), then this will be evidence for a
violation of CPT.
\vskip 0.1in
\noindent
{\it Acknowledgements}: 
We thank S. L. Glashow, S. Weinberg, and E. Witten for discussions. 
This research was supported by the NSF under 
Grant No. PHY0070928 and the DOE  
under Grants No.~DE-FG02-95ER40896 and No.~DE-FG02-91ER40676  
and by the WARF.

\end{document}